# Novel Evolution of the Mineralocorticoid Receptor in Humans compared to Chimpanzees, Gorillas and Orangutans

Yoshinao Katsu[1,2,] *, Jiawen Zhang[2], Michael E. Baker[3,4,] *

[1] Faculty of Science

Hokkaido University

Sapporo, Japan

[2] Graduate School of Life Science

Hokkaido University

Sapporo, Japan

[3] Division of Nephrology-Hypertension

Department of Medicine, 0693

University of California, San Diego

9500 Gilman Drive

La Jolla, CA

Center for Academic Research and Training in Anthropogeny (CARTA) [4]

University of California, San Diego

La Jolla, CA

*Correspondence to

Y. Katsu; E-mail: ykatsu@sci.hokudai.ac.jp (YK)

M. E. Baker; E-mail: mbaker@health.ucsd.edu (MEB)

**Abstract**

Five distinct full-length mineralocorticoid receptor (MR) genes have been identified in humans. These human MRs can be distinguished by the presence or absence of an in-frame insertion of 12 base pairs coding for Lys, Cys, Ser, Trp (KCSW) in their DNA-binding domain (DBD) and the presence of two amino acid mutations in their amino terminal domain (NTD). Two human MRs with the KCSW insertion (MR-KCSW) and three human MRs without KCSW in the DBD have been identified. The three human MRs without KCSW contain either (Ile-180, Ala-241) or (Val-180, Val-241) or (Ile-180, Val-241) in their NTD. The two human MRs with KCSW contain either (Val-180, Val-241) or (Ile-180, Val-241) in their NTD. Human MR-KCSW with (Ile-180, Ala-241) has not been cloned. In contrast, chimpanzees contain two MRs with KCSW and two MRs without KCSW in their DBD and both contain only Ile180, Val-241 in their NTDs. Each pair of chimpanzee MRs differ at another amino acid in the NTD. A chimpanzee MR with either Val-180, Val-241 or Ile-180, Ala-241 in the NTD has not been cloned. Gorillas and orangutans each contain one MR with KCSW in the DBD and one MR without KCSW. Both gorilla and orangutan MRs contain I-180, Val-241 in their NTD. Neither Val-180, Val-241 nor Ile-180, Ala-241 are found in the NTD in either a gorilla MR or an orangutan MR. These data suggest that human MRs with Val-180, Val-241 or Ile-180, Ala-241 in the NTD evolved after humans and chimpanzees diverged from their common ancestor. These unique human MRs may have had a role in the divergent evolution of humans from chimpanzees. Studies are underway to characterize transcriptional activation of the five human MRs by aldosterone, cortisol, and other corticosteroids for comparison with each other to elucidate the roles of these MRs in human physiology.

Key words: Mineralocorticoid Receptor, DNA-binding domain, Evolution, Human MR, Chimpanzee MR, Aldosterone

Running title: Evolution of the DNA-binding domain of the Mineralocorticoid Receptor

## Introduction

The mineralocorticoid receptor (MR) is a ligand-activated transcription factor, belonging to the nuclear receptor family, a diverse group of transcription factors that arose in multicellular animals [1–5]. The traditional physiological function of the MR is to maintain electrolyte balance by regulating sodium and potassium transport in epithelial cells in the kidney and colon [6–10]. In addition, the MR has important physiological functions in many other tissues, including brain, heart, skin and lungs [10–16].

The MR and its paralog, the glucocorticoid receptor (GR), descended from an ancestral corticoid receptor (CR) in a cyclostome (jawless fish) that evolved about 550 million years ago at the base of the vertebrate line [17–24]. A descendent of this ancestral steroid receptor, the CR in lamprey (*Petromyzon marinus*), is activated by aldosterone [25,26] and other corticosteroids [25,26]. Lampreys contain two CR isoforms, which differ only in the presence of a four amino acid insert Thr, Arg, Gln, Gly (TRQG) in their DNA-binding domain (DBD) [27] (Figure 1). We found that several corticosteroids had a similar half-maximal response (EC50) for lamprey CR1 and CR2 [26]. However, these corticosteroids had a lower fold-activation of transcription for CR1, which contains the four amino acid insert, than for CR2 suggesting that the deletion of the four amino acid sequence in CR2 selected for increased transcriptional activation by corticosteroids of CR2 [26,27].

A distinct MR and GR first appear in sharks and other cartilaginous fishes (Chondrichthyes) [19,21,28–31]. The DBD in elephant shark MR and GR lacks the four amino acid sequence found in lamprey CR1 [26] (Figure 1). We inserted this four-residue sequence from lamprey CR1 into the DBD in elephant shark MR and GR and found that in HEK293 cells co-transfected with the TAT3 promoter, the mutant elephant shark MR and GR had lower transcriptional activation by corticosteroids than did their wild-type elephant shark MR and GR counterparts, indicating that the insertion of the four amino acid sequence into the DBD of wild-type elephant shark MR and GR had a similar effect on transcriptional activation as the KCSW insert had in the DBD of lamprey CR1 [27].

We then analyzed the DBD sequence of human MR, which had been cloned, sequenced and characterized by Arriza et al. [32], and found that like elephant shark MR, this human MR (MR1) lacks a four-residue segment in its DBD. This human MR has been widely studied [8,10,12,13,33–35]. Unexpectedly, our BLAST [36] search with the DBD from this human MR found a second, previously described, human MR splice variant with a KCSW insert (MR-KCSW) in its DBD [27,37–39] (Figure 1). As described later, further BLAST searches found two full-length human MRs with this insert (MR-KCSW) and three full-length human MRs without this insert. The three human MRs without the KCSW insert contain either (Ile-180, Ala-241) or (Val-180, Val-241) or (Ile-180, Val-241) in their amino terminal domain (NTD). The two human MR-KCSW splice variants contain either (Val-180, Val-241) or (Ile-180, Val-241) in their NTD. A human MR-KCSW with (Ile-180, Ala-241) has not been cloned.

Although the level of expression of MR-KCSW in human and rat tissues has been reported [37–39], transcriptional activation of either human or rat MR-KCSW by corticosteroids has not been reported. To remedy this deficiency, we have studies underway to characterize transcriptional activation of the five full-length human MRs by a panel of corticosteroids. Here, in this brief report, we describe our evolutionary analysis of human MR from a comparison of five full-length human MRs with four full-length MRs in chimpanzees and two full-length MRsin gorillas, and orangutans. Due to the multiple functions in human development of the MR alone [11,15,16,40], as well as due to the MRs interaction with the GR [41–45], we suggest that the sequence divergence of human MR from chimpanzee MR may have been important in the evolution of humans from chimpanzees.

```
Human MR1:          CLVCGDEASGCHYGVVTCGSCKVFFKRAVEG----QHNYLCAGRNDCIIDKIRRKNCPACRLQKCLQAGM
Human MR-KCSW:      CLVCGDEASGCHYGVVTCGSCKVFFKRAVEGKCSWQHNYLCAGRNDCIIDKIRRKNCPACRLQKCLQAGM
Rat MR:             CLVCGDEASGCHYGVVTCGSCKVFFKRAVEG----QHNYLCAGRNDCIIDKIRRKNCPACRLQKCLQAGM
Rat MR-KCSW:        CLVCGDEASGCHYGVVTCGSCKVFFKRAVEGKCSWQHNYLCAGRNDCIIDKIRRKNCPACRLQKCLQAGM
Platypus MR:        CLVCGDEASGCHYGVVTCGSCKVFFKRAVEG----QHNYLCAGRNDCIIDKIRRKNCPACRLQKCLQAGM
Platypus MR-KCSW:   CLVCGDEASGCHYGVVTCGSCKVFFKRAVEGKCSWQHNYLCAGRNDCIIDKIRRKNCPACRLQKCLQAGM
Turtle MR:          CLVCGDEASGCHYGVVTCGSCKVFFKRAVEG----QHNYLCAGRNDCIIDKIRRKNCPACRLQKCLQAGM
Turtle MR-KCSW:     CLVCGDEASGCHYGVVTCGSCKVFFKRAVEGKCSWQHNYLCAGRNDCIIDKIRRKNCPACRLQKCLQAGM
Xenopus MR:         CLVCGDEASGCHYGVVTCGSCKVFFKRAVEG----QHSYLCAGRNDCIIDKIRRKNCPACRLQKCLQAGM
Xenopus MR-KCSR:    CLVCGDEASGCHYGVVTCGSCKVFFKRAVEGKCSRQHSYLCAGRNDCIIDKIRRKNCPACRLQKCLQAGM
Lungfish MR:        CLVCGDEASGCHYGVVTCGSCKVFFKRAVEG----QHNYLCAGRNDCIIDKIRRKNCPACRVRKCLQAGM
Zebrafish MR:       CLVCGDEASGCHYGVVTCGSCKVFFKRAVEG----QHNYLCAGRNDCIIDKIRRKNCPACRVRKCLQAGM
Elephant shark MR:CLVCSDEASGCHYGVLTCGSCKVFFKRAVEG----QQNYLCAGRNDCIIDKIRRKNCPACRLRKCLKAGM
Lamprey CR1:        CLICSDEASGCHYGVLTCGSCKVFFKRAVEGTRQGQHNYLCAGRNDCIIDKIRRKNCPACRLRKCIQAGM
Lamprey CR2:        CLICSDEASGCHYGVLTCGSCKVFFKRAVEG----QHNYLCAGRNDCIIDKIRRKNCPACRLRKCIQAGM
```

**Figure 1. Comparison of the DNA-binding domain on human MR1, human MR-KCSW, rat MR, rat MR-KCSW, Platypus MR, Platypus MR-KCSW, Turtle MR, Turtle MR-KCSW, Xenopus MR, Xenopus MR-KCSR, lungfish MR, zebrafish MR, elephant shark MR, and lamprey CR1 and CR2.**

**Figure legend:** The DBD of the human MR-KCSW splice variant has an insertion of four amino acids that is absent in human MR1. Otherwise, the rest of the sequences of human MR and human MR-KCSW are identical. Moreover, except for the four amino acid insert in the DBD of human MR, the DBDs of rat MR, platypus MR and turtle MR are identical. The four amino acid splice variant in the DBD of *Xenopus* MR differs from these MRs by one amino acid. Differences between the DBD sequence in human MR and selected vertebrate MRs are shown in red. Protein accession numbers are AAA59571 for human MR, XP_011530277 for human MR-

KCSW, NP_037263 for rat MR, XP_038953451 for rat MR-KCSW, XP_007669969 for platypus MR, XP_016083764 for platypus MR-KCSW, XP_043401725 for turtle MR, XP_037753298 for turtle MR-KCSW, NP_001084074 for *Xenopus* MR, XP_018098693 for *Xenopus* MR-KCSR, BCV19931 for lungfish MR, NP_001093873 for zebrafish MR, XP_007902220 for elephant shark MR, XP_032811370 for lamprey CR1, and XP_032811371 for lamprey CR2.

## Results and Discussion

### Humans have five full-length MRs

A BLAST [36] search of GenBank retrieved five distinct full-length human MR genes (Figure 2). We retrieved three full-length human MRs without a KCSW insert in the DBD. These three human MRs contain either (Ile-180, Ala-241), (Val-180, Val-241) or (Ile-180, Val-241) in the NTD (Figure 2). We also retrieved two human MR-KCSWs containing either (Val-180, Val-241) or (Ile-180, Val-241) in their NTD. A human MR-KCSW sequence with (Ile-180, Ala-241) has not been deposited in GenBank.

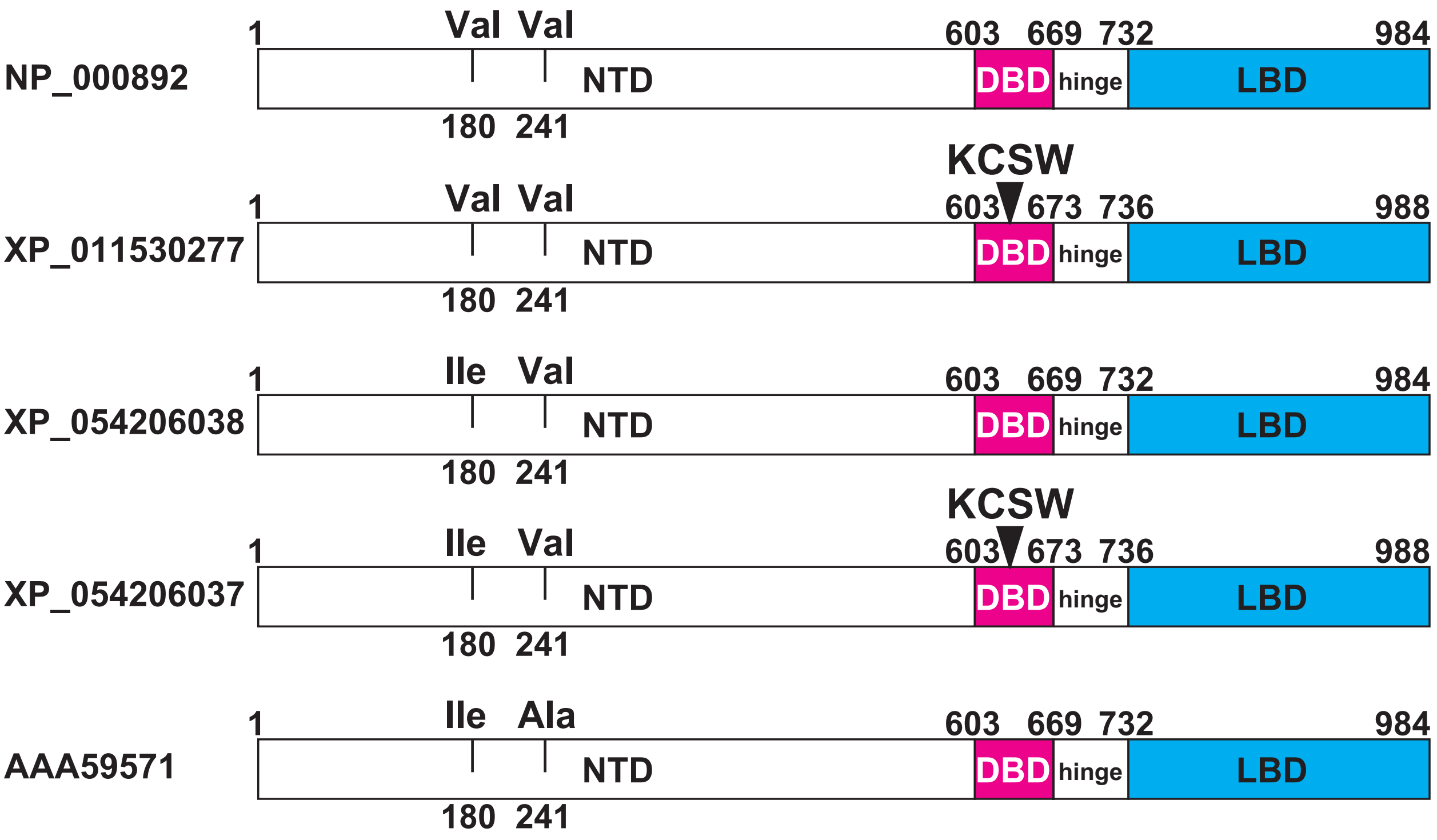

**Figure 2. Comparison of five full-length human MRs.**

**Figure legend:** There are three human MRs with 984 amino acids. They contain either valine-180 and valine-241, isoleucine-180 and valine-241 or isoleucine-180 and alanine-241 in the amino-terminal domain (NTD). There are two human MRs with 988 amino acids and KCSW in the DBD and either valine-180 and valine-241 or isoleucine-180 and valine-241 in the NTD. A human MR with isoleucine-180 and alanine-241 and a KCSW insert in the NTD was not found in GenBank.

## Chimpanzees have four full-length MRs

Chimpanzees contain two full-length MRs that have DBDs that are identical to the DBD in human MRs without the KCSW insert and two MRs that have DBDs that are identical to the DBD in human MR-KCSW (Figure 1, Figure 3). All four chimpanzee MRs contain (Ile-180, Val-241) in their NTD, which is similar to the NTD in one human MR (Figure 2). Chimpanzees lack an MR with either (Ile-180, Ala-241) or (Val-180, Val-241) in their NTD. The different chimpanzee MR sequences contain a Ser-591 to Asn-591 mutation in their NTD (Figure 3).

Human MRs contain a serine-591 corresponding to serine-591 in chimpanzee MR.

```
MR1:       METKGYHSLPEGLDMERRWGQVSQAVEHSSLGPTERTDENNYMEIVNVSCVSGAIPNNSTQGSSKEKHELLPCLQQDNNRPGILTSDIKT  90
MR2-KCSW:  METKGYHSLPEGLDMERRWGQVSQAVEHSSLGPTERTDENNYMEIVNVSCVSGAIPNNSTQGSSKEKHELLPCLQQDNNRPGILTSDIKT  90
MR3:       METKGYHSLPEGLDMERRWGQVSQAVEHSSLGPTERTDENNYMEIVNVSCVSGAIPNNSTQGSSKEKHELLPCLQQDNNRPGILTSDIKT  90
MR4-KCSW:  METKGYHSLPEGLDMERRWGQVSQAVEHSSLGPTERTDENNYMEIVNVSCVSGAIPNNSTQGSSKEKHELLPCLQQDNNRPGILTSDIKT  90
           ******************************************************************************************

MR1:       ELESKELSATVAESMGLYMDSVRDADYSYEQQNQQGSMSPAKIYQNVEQLVKFYKGNGHRPSTLSCVNRPLRSFMSDSGSSVNGGVMRAI 180
MR2-KCSW:  ELESKELSATVAESMGLYMDSVRDADYSYEQQNQQGSMSPAKIYQNVEQLVKFYKGNGHRPSTLSCVNRPLRSFMSDSGSSVNGGVMRAI 180
MR3:       ELESKELSATVAESMGLYMDSVRDADYSYEQQNQQGSMSPAKIYQNVEQLVKFYKGNGHRPSTLSCVNRPLRSFMSDSGSSVNGGVMRAI 180
MR4-KCSW:  ELESKELSATVAESMGLYMDSVRDADYSYEQQNQQGSMSPAKIYQNVEQLVKFYKGNGHRPSTLSCVNRPLRSFMSDSGSSVNGGVMRAI 180
           ******************************************************************************************

MR1:       VKSPIMCHEKSPSVCSPLNMTSSVCSPAGINSVSSTTASFGNFPVHSPITQGTPLTCSPNVENRGSRSHSPAHASNVGSPLSSPLSSMKS 270
MR2-KCSW:  VKSPIMCHEKSPSVCSPLNMTSSVCSPAGINSVSSTTASFGNFPVHSPITQGTPLTCSPNVENRGSRSHSPAHASNVGSPLSSPLSSMKS 270
MR3:       VKSPIMCHEKSPSVCSPLNMTSSVCSPAGINSVSSTTASFGNFPVHSPITQGTPLTCSPNVENRGSRSHSPAHASNVGSPLSSPLSSMKS 270
MR4-KCSW:  VKSPIMCHEKSPSVCSPLNMTSSVCSPAGINSVSSTTASFGNFPVHSPITQGTPLTCSPNVENRGSRSHSPAHASNVGSPLSSPLSSMKS 270
           ******************************************************************************************

MR1:       SISSPPSHCSVKSPVSSPNNVTLRSSVSSPANINNSRCSVSSPSNTNNRSTLSSPAASTVGSICSPVNNAFSYTASGTSAGSSTLRDVVP 360
MR2-KCSW:  SISSPPSHCSVKSPVSSPNNVTLRSSVSSPANINNSRCSVSSPSNTNNRSTLSSPAASTVGSICSPVNNAFSYTASGTSAGSSTLRDVVP 360
MR3:       SISSPPSHCSVKSPVSSPNNVTLRSSVSSPANINNSRCSVSSPSNTNNRSTLSSPAASTVGSICSPVNNAFSYTASGTSAGSSTLRDVVP 360
MR4-KCSW:  SISSPPSHCSVKSPVSSPNNVTLRSSVSSPANINNSRCSVSSPSNTNNRSTLSSPAASTVGSICSPVNNAFSYTASGTSAGSSTLRDVVP 360
           ******************************************************************************************

MR1:       SPDTQEKGAQEVPFPKTEEVESAISNGVTGQLNIVQYIKPEPDGAFSSSCLGGNSKINSDSSFSVPIKQESTKHSCSGTSFKGNPTVNPF 450
MR2-KCSW:  SPDTQEKGAQEVPFPKTEEVESAISNGVTGQLNIVQYIKPEPDGAFSSSCLGGNSKINSDSSFSVPIKQESTKHSCSGTSFKGNPTVNPF 450
MR3:       SPDTQEKGAQEVPFPKTEEVESAISNGVTGQLNIVQYIKPEPDGAFSSSCLGGNSKINSDSSFSVPIKQESTKHSCSGTSFKGNPTVNPF 450
MR4-KCSW:  SPDTQEKGAQEVPFPKTEEVESAISNGVTGQLNIVQYIKPEPDGAFSSSCLGGNSKINSDSSFSVPIKQESTKHSCSGTSFKGNPTVNPF 450
           ******************************************************************************************

MR1:       PFMDGSYFSFMDDKDYYSLSGILGPPVPGFDGNCEGSGFPVGIKQEPDDGSYYPEASIPSSAIVGVNSGGQSFHYRIGAQGTISLSRSAR 540
MR2-KCSW:  PFMDGSYFSFMDDKDYYSLSGILGPPVPGFDGNCEGSGFPVGIKQEPDDGSYYPEASIPSSAIVGVNSGGQSFHYRIGAQGTISLSRSAR 540
MR3:       PFMDGSYFSFMDDKDYYSLSGILGPPVPGFDGNCEGSGFPVGIKQEPDDGSYYPEASIPSSAIVGVNSGGQSFHYRIGAQGTISLSRSAR 540
MR4-KCSW:  PFMDGSYFSFMDDKDYYSLSGILGPPVPGFDGNCEGSGFPVGIKQEPDDGSYYPEASIPSSAIVGVNSGGQSFHYRIGAQGTISLSRSAR 540
           ******************************************************************************************

MR1:       DQSFQHLSSFPPVNTLVESWKSHGDLSSRRSDGYPVLEYIPENVSSSTLRSVSTGSSRPSKICLVCGDEASGCHYGVVTCGSCKVFFKRA 630
MR2-KCSW:  DQSFQHLSSFPPVNTLVESWKSHGDLSSRRSDGYPVLEYIPENVSSSTLRSVSTGSSRPSKICLVCGDEASGCHYGVVTCGSCKVFFKRA 630
MR3:       DQSFQHLSSFPPVNTLVESWKSHGDLSSRRSDGYPVLEYIPENVSSSTLRNVSTGSSRPSKICLVCGDEASGCHYGVVTCGSCKVFFKRA 630
MR4-KCSW:  DQSFQHLSSFPPVNTLVESWKSHGDLSSRRSDGYPVLEYIPENVSSSTLRNVSTGSSRPSKICLVCGDEASGCHYGVVTCGSCKVFFKRA 630
           ************************************************** ***************************************

MR1:       VEG----QHNYLCAGRNDCIIDKIRRKNCPACRLQKCLQAGMNLGARKSKKLGKLKGIHEEQPQQQQPPPPPPPPQSPEEGTTYIAPAKE 716
MR2-KCSW:  VEGKCSWQHNYLCAGRNDCIIDKIRRKNCPACRLQKCLQAGMNLGARKSKKLGKLKGIHEEQPQQQQPPPPPPPPQSPEEGTTYIAPAKE 720
MR3:       VEG----QHNYLCAGRNDCIIDKIRRKNCPACRLQKCLQAGMNLGARKSKKLGKLKGIHEEQPQQQQPPPPPPPPQSPEEGTTYIAPAKE 716
MR4-KCSW:  VEGKCSWQHNYLCAGRNDCIIDKIRRKNCPACRLQKCLQAGMNLGARKSKKLGKLKGIHEEQPQQQQPPPPPPPPQSPEEGTTYIAPAKE 720
           ***    ***********************************************************************************

MR1:       PSVNTALVPQLSTISRALTPSPVMVLENIEPEIVYAGYDSSKPDTAENLLSTLNRLAGKQMIQVVKWAKVLPGFKNLPLEDQITLIQYSW 806
MR2-KCSW:  PSVNTALVPQLSTISRALTPSPVMVLENIEPEIVYAGYDSSKPDTAENLLSTLNRLAGKQMIQVVKWAKVLPGFKNLPLEDQITLIQYSW 810
MR3:       PSVNTALVPQLSTISRALTPSPVMVLENIEPEIVYAGYDSSKPDTAENLLSTLNRLAGKQMIQVVKWAKVLPGFKNLPLEDQITLIQYSW 806
MR4-KCSW:  PSVNTALVPQLSTISRALTPSPVMVLENIEPEIVYAGYDSSKPDTAENLLSTLNRLAGKQMIQVVKWAKVLPGFKNLPLEDQITLIQYSW 810
           ******************************************************************************************

MR1:       MCLSSFALSWRSYKHTNSQFLYFAPDLVFNEEKMHQSAMYELCQGMHQISLQFVRLQLTFEEYTIMKVLLLLSTIPKDGLKSQAAFEEMR 896
MR2-KCSW:  MCLSSFALSWRSYKHTNSQFLYFAPDLVFNEEKMHQSAMYELCQGMHQISLQFVRLQLTFEEYTIMKVLLLLSTIPKDGLKSQAAFEEMR 900
MR3:       MCLSSFALSWRSYKHTNSQFLYFAPDLVFNEEKMHQSAMYELCQGMHQISLQFVRLQLTFEEYTIMKVLLLLSTIPKDGLKSQAAFEEMR 896
MR4-KCSW:  MCLSSFALSWRSYKHTNSQFLYFAPDLVFNEEKMHQSAMYELCQGMHQISLQFVRLQLTFEEYTIMKVLLLLSTIPKDGLKSQAAFEEMR 900
           ******************************************************************************************

MR1:       TNYIKELRKMVTKCPNNSGQSWQRFYQLTKLLDSMHDLVSDLLEFCFYTFRESHALKVEFPAMLVEIISDQLPKVESGNAKPLYFHRK 984
MR2-KCSW:  TNYIKELRKMVTKCPNNSGQSWQRFYQLTKLLDSMHDLVSDLLEFCFYTFRESHALKVEFPAMLVEIISDQLPKVESGNAKPLYFHRK 988
MR3:       TNYIKELRKMVTKCPNNSGQSWQRFYQLTKLLDSMHDLVSDLLEFCFYTFRESHALKVEFPAMLVEIISDQLPKVESGNAKPLYFHRK 984
MR4-KCSW:  TNYIKELRKMVTKCPNNSGQSWQRFYQLTKLLDSMHDLVSDLLEFCFYTFRESHALKVEFPAMLVEIISDQLPKVESGNAKPLYFHRK 988
           ****************************************************************************************
```

**Figure 3. Multiple alignment of four chimpanzee MRs.**

**Figure legend:** The four chimpanzee MRs were aligned with Clustal Omega [46]. MR1 and MR2-KCSW have serine-591 and MR3 and MR4-KCSW have asparagine-591 in the NTD.

## Gorillas and orangutans have two full-length MRs

Gorillas and orangutans each contain two full-length MRs: one MR with 984 amino acids and one MR with 988 amino acids. The MRs with 988 amino acids have the KCSW sequence in the DBD. All full-length gorilla MRs and orangutan MRs have isoleucine-180 and valine-241 in the amino-terminal domain (NTD) (Figure 4). A gorilla MR or orangutan MR with either valine-180 and valine-241 or isoleucine-180 and alanine-241 in the NTD was not found in GenBank.

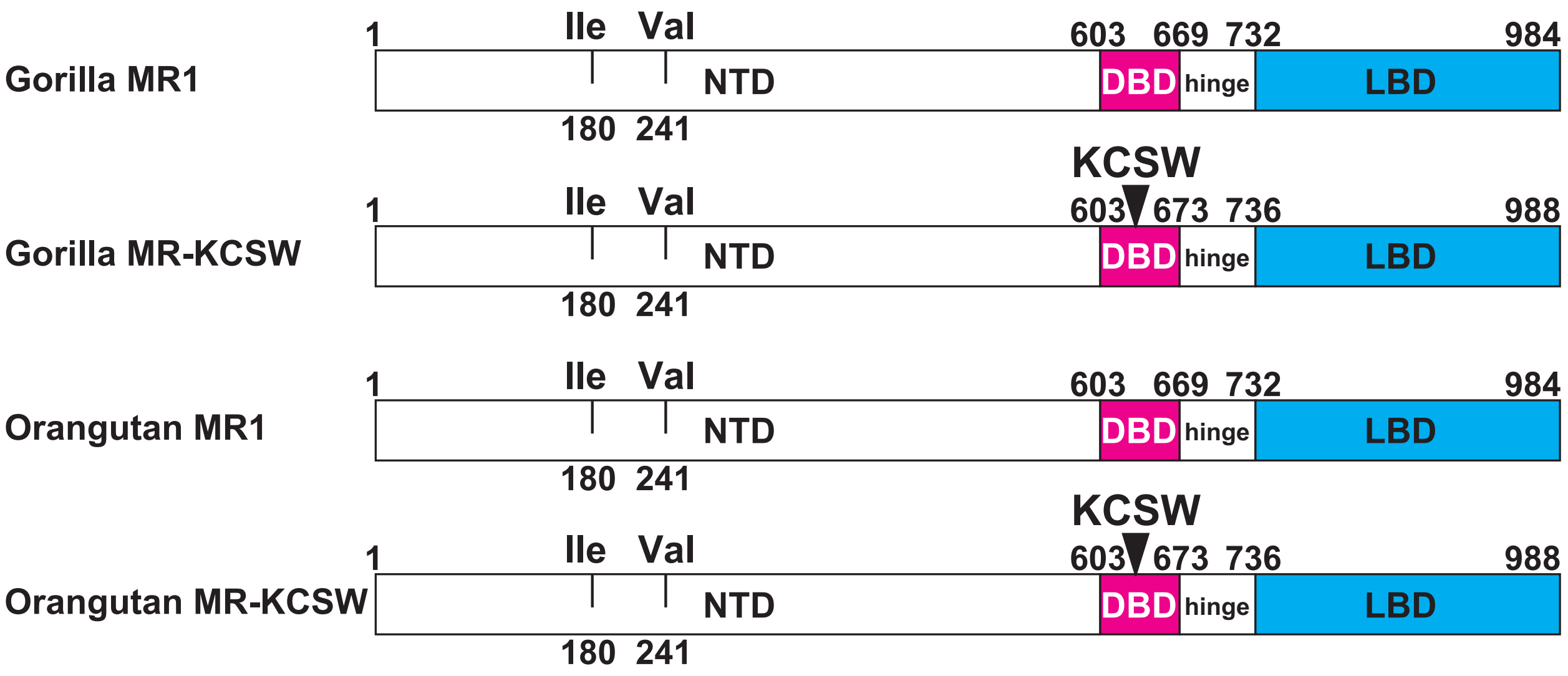

**Figure 4. Comparison of gorilla and orangutan MR sequences.**

**Figure legend:** Gorilla MRs and Orangutan MRs conserve an isoleucine-180, valine-241 pair corresponding to isoleucine-180 and valine-241 in the NTD of human MR. A gorilla MR or an orangutan MR with either valine-180 and valine-241 or isoleucine-180 and alanine-241 in the NTD was not found in GenBank.

## Evolutionary divergence of human and chimpanzee MRs

Our analysis of human and chimpanzee MRs indicates that human MR with either (Ile-180, Ala-241) or (Val-180, Val-241) in the NTD evolved after the divergence of humans and chimpanzees from a common ancestor. The physiological consequences of (Ile-180, Ala-241) or (Val-180, Val-241) in the NTD of human MR remain to be elucidated. As a first step, to determine if there are differences in transcriptional activation of these MRs by corticosteroids, we have cloned these human MR sequences and are screening them for transcriptional activation by aldosterone, cortisol, and other corticosteroids.

**Evolution of the MR DBD in basal terrestrial vertebrates**

The evolution of an MR with a DBD containing a four amino acid insert in human MR and other terrestrial MRs was surprising to us (Figure 1). We did not expect to find the KCSW insert in human MR at the position homologous to the position of TRQG in the DBD of lamprey CR because an insert at this position was not present in the DBD of elephant sharks and lungfish MRs (Figure 1). Moreover, the DBD sequence is highly conserved in terrestrial vertebrates (Figure 1) suggesting that the evolution of this four amino acid sequence in the DBD has an important function in terrestrial vertebrates. The evolution of the KCSR sequence into the DBD *Xenopus* MR (Figure 1) places the re-emergence of this motif close to the origin of terrestrial vertebrates. Moreover, turtles contain an MR with KCSW in the homologous position in DBD (Figure 1) of human MR DBD, indicating that KCSW also has an ancient origin in terrestrial vertebrate MRs.

**Funding:** This work was supported by Grants-in-Aid for Scientific Research from the Ministry of Education, Culture, Sports, Science and Technology of Japan (23K05839) to Y.K., and the Takeda Science Foundation to Y.K.

**Competing interests:** The authors have declared that no competing interests exist.

**Contributions**.

Yoshinao Katsu: Investigation, Conceptualization, Supervision, Formal Analysis, Writing – original draft, Writing – review & editing.

Jiawen Zhan: Data curation, Investigation, Methodology.

Michael E. Baker: Conceptualization; Formal analysis, Supervision, Writing – original draft, Writing – review & editing.